# Optical Bar Code for Internet Access Application based on Optical camera communication and Bluetooth Control

**Shangsheng Wen**[1*], **Manxi Liu**[2], **Yanyi Chen**[2], **Yirong Chen**[3], **Futong An**[3], **Yingcong Chen**[4], **Weipeng Guan**[2]

[1] *School of Materials Science and Engineering, South China University of Technology, Guangzhou , Guangdong 510640, China*
[2] *School of Automation Science and Engineering, South China University of Technology, Guangzhou, Guangdong 510640, China*
[3] *School of Electronic and Information Engineering, South China University of Technology, Guangzhou, Guangdong 510640, China*
[4]*Foshan Wengu Technology Co.,Ltd, Foshan, Guangdong 510640, China*
*\* Correspondence: shshwen@scut.edu.cn*

**Abstract:** We demonstrate an internet access application based on optical camera communication and bluetooth. The app will access the website while the camera in the phone receives the optical signal. © 2022 The Author(s)

## 1. Overview

Due to insufficient spectrum resources of traditional radio frequency (RF) , visible light communication (VLC) has proposed an effective solution. As a technical solution of VLC, Optical Camera Communication(OCC) shows broad application prospects due to the wide distribution of smart phone cameras. It uses complementary metal oxide semiconductor (CMOS) image-sensor to captures the LED image received by the camera. There are generally two ways to capture video signals, one is global shutter(GS) exposure, and the other is rolling shutter exposure(RSE)[1]. All the pixels of the former are exposed at the same time, and the data transfer rate is severely limited by the frame rate. Pixels in the latter are exposed row by row, this phenomenon is called rolling curtain effect. Compared to GS, RSE can detect multiple states in the same frame. Therefore, its data rate is much higher than video frame rate, which helps to improve the performance of the communication system. Instead of modifying the hardware at the receiving end of a smartphone, users can use the camera to extract a rich array of information from the LED by embedding a tiny chip into it. Meanwhile, OCC provides efficient solutions for Internet of Things(IOT) fields, and has attracted widespread attention. [2] proposes a multi-sensor fusion method based on VLP and SLAM to provide reliable location in LED outage situations. [3] proposes an indoor robot VLP localization system based on ROS, which achieves high positioning accuracy and low calculation time. [4] demonstrate a OCC system that allowed the user's smartphone to decode real-time information from the optical signal reflected by the poster's surface. [5] propose an underwater optical communication(UWOC).

In this demonstration, we proposeed an internet access application based on optical bar code. The application presented that the smart phone successfully captured and decoded the information sent by the transmitter, and then access to the corresponding website according to the information. Without storing the data in the control circuit module in advance, users can obtain a wealth of changeable information from the optical bar code of the LED. This technology has huge application potential, such as: exhibition narration and introduction, conference check-in, advertising and product information acquisition, etc.

## 2. Innovation

The main contributions of the demonstration are as follows:

(1)Design a LED driver modulation module based on bluetooth control. The system uses the bluetooth module for wireless data passthrough. Meanwhile, on-off-key(OOK) modulation is used to control the on-off state of LED. The control terminal can modify the optical signal instructions sent by LED through bluetooth module.

(2)Design a internet access application based on optical bar code. Not only can it use the image processing algorithm to filter and decode the received image, but also the decoded data and optical bar code will be displayed on the app interface and access to the corresponding website.

(3)Based on the above designs, we demonstrate an OCC experimental platform based on Bluetooth control. When the front camera of the mobile phone receives the optical signal, it will display the decoding result and optical bar code in the app and access to the corresponding website.

## 3. Description of Demonstration

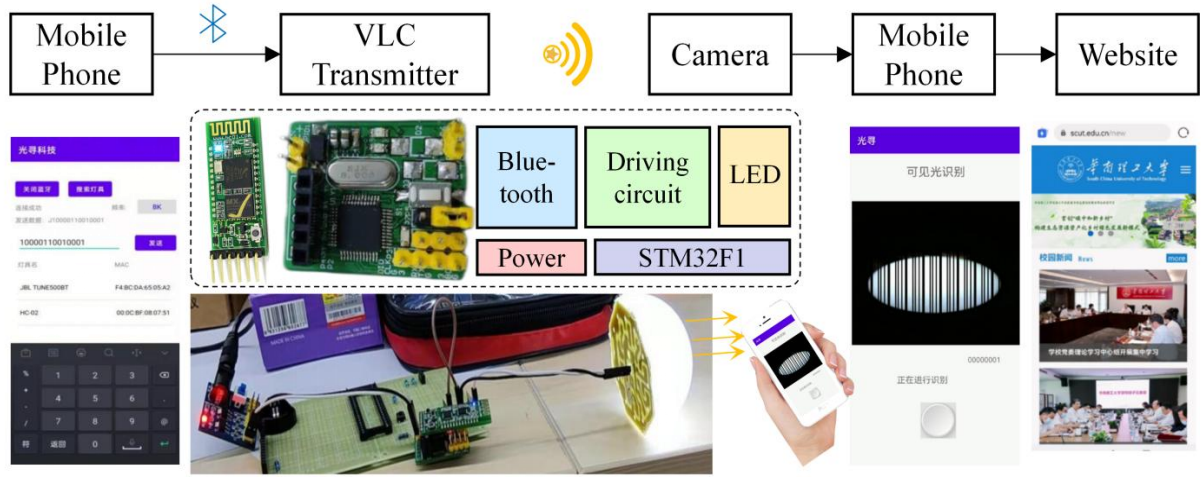

Fig. 1. Demonstration setup

In the experimental platform, In our experimental platform, the VLC transmitter includes: bluetooth, driving circuit, LED, STM32F1 and power. First of all, the 220V AC power is transformed into a 5V DC power through a voltage regulator module. Subsequently, 5V DC power drives the LED and driving circuit, on the other hand, it is transformed to a 3.3V DC power by ams117 voltage regulator module to drive the bluetooth, STM32F1, and driving circuit modules.

After connecting to bluetooth, We sends data to HC-02 Bluetooth Low Energy SoC integrated memory through our remote app, then the data is transferred and loaded to the single port read-only memory (ROM) of STM32F1. The microcontroller generates a modulation signal based on OOK method and outputs high and low voltage to the input port of the driving circuit(IPDC) through the general-purpose input/output (GPIO). The IPDC is not directly driving the MOSFET, but an NPN Bipolar Junction Transistor (BJT) is inserted between the IPDC and the gate of the MOSFET to raise the gate voltage of the MOSFET to make it larger than the open Voltage $U_{gs\,(th)}$, while reducing the heat caused by high-frequency switching. At the same time, the LED is in the on-off state according to the the voltage of the IPDC to realize the transmission of the optical signal. On the receiving end, we placed a Huawei Nova5 Pro smartphone horizontally on the desktop. Then we picked up the LED and mounted it on top of the phone so the front-facing camera could receive the optical signal from the VLC transmitter. During the recording process, the information transmitted between frames is lost because it cannot be captured by the camera. Therefore, in order to ensure the reliable transmission of data, we encapsulate the data into a data packet for transmission. It is composed of a fixed header 100001 and 8-bit data, and the continuous bit 0 of the data part should be less than 3 bits to facilitate the identification and recovery of the data. At the decoding end, the OCC application was developed with Java language through Andriod studio. Meanwhile, we use Huawei Nova5 Pro smartphones, and finetune the camera parameters to achieve the best decoding effect.

For the rolling shutter effect, a series of black and white stripes are generated in the LED image, then image processing algorithms are used to filter and decode the received image. As long as we click the "Start recognition" button in the app, the app will display the optical code stripe and the decoded bit sequence, and the decoded bit sequence can also be converted to control instructions, so as to access to websites at the receiving end. The video record of our demo can be seen in the website[1][2][3]. More information about the OCC's work can be found at:[6]-[9]. More work on VLP can be found at:[10]-[12].

## 4. OFC relevance

As an important optical communication technology, OCC is a hot trending of OFC. This demonstration involves an optical Bar Code for Internet Access Application based on Optical camera communication and Bluetooth Control. The LED is controlled by our app and driving circuit, allowing users to detect optical signals through the camera in the mobile phone to access to different websites.

---

[1] https://b23.tv/5SMp8qV
[2] https://b23.tv/N7P5bo5
[3] https://b23.tv/f1QzV2C